\def\year{2021}\relax
\documentclass[letterpaper]{article} 
\usepackage{aaai21}  
\usepackage{times}  
\usepackage{helvet} 
\usepackage{courier}  
\usepackage[hyphens]{url}  
\usepackage{graphicx} 
\usepackage{natbib}  
\usepackage{caption} 
\usepackage[draft]{hyperref}
\usepackage{graphicx,balance,multirow,xspace,subcaption,longtable}
\usepackage{makecell,amsmath,cleveref,rotating,amssymb}
\usepackage[font={small}, textfont=md]{caption}
\usepackage[T1]{fontenc}
\usepackage[draft,inline,nomargin,index]{fixme}
\usepackage{booktabs}
\usepackage{xcolor}
\usepackage[utf8]{inputenc}

\PassOptionsToPackage{hyphens}{url}
\newcolumntype{P}[1]{>{\arraybackslash}p{#1}}
\newcolumntype{X}[1]{>{\centering\arraybackslash}p{#1}}

\expandafter\def\expandafter\UrlBreaks\expandafter{\UrlBreaks
  \do\a\do\b\do\c\do\d\do\e\do\f\do\g\do\h\do\i\do\j%
  \do\k\do\l\do\m\do\n\do\o\do\p\do\q\do\r\do\s\do\t%
  \do\u\do\v\do\w\do\x\do\y\do\z\do\A\do\B\do\C\do\D%
  \do\E\do\F\do\G\do\H\do\I\do\J\do\K\do\L\do\M\do\N%
  \do\O\do\P\do\Q\do\R\do\S\do\T\do\U\do\V\do\W\do\X%
  \do\Y\do\Z}

\crefformat{section}{\S#2#1#3}
\crefformat{subsection}{\S#2#1#3}
\crefformat{subsubsection}{\S#2#1#3}

\newcommand\clearrow{\global\let\rowmac\relax}

\newcommand{\para}[1]{{\vspace{.05in} \bf \noindent #1 }}
\newcommand{\parait}[1]{{\vspace{.05in} \em \noindent #1 }}

\newcommand{\subreddit}[1]{\emph{r/#1}}

\newcommand{\D}[1]{$\mathfrak{D}_{#1}$}

\renewcommand{\footnotesize}{\scriptsize}

\newcommand{\eg}{e.g.,\ }
\newcommand{\etal}{et al.\xspace}
\newcommand{\ie}{i.e.,\ }

\title{Reddit and the Fourth Estate: \\ Exploring the magnitude and effects of media influence on community level moderation on Reddit}

\author {
     Hussam Habib,
     Rishab Nithyanand \\
 }
 \affiliations {
     University of Iowa \\
     hussam-habib@uiowa.edu, rishab-nithyanand@uiowa.edu
 }

\begin{document}

\maketitle

\sloppy
\begin{abstract}

  Most platforms, including Reddit, face a dilemma when applying interventions
  such as subreddit bans to toxic communities --- do they risk angering their user
  base by proactively enforcing stricter controls on discourse or do they defer
  interventions at the risk of eventually triggering negative media reactions
  which might impact their advertising revenue? 
  In this paper, we analyze Reddit's previous administrative interventions to
  understand one aspect of this dilemma: the relationship between the media and
  administrative interventions. More specifically, we make two primary
  contributions. First, using a mediation analysis framework, we find evidence
  that Reddit's interventions for violating their content policy for toxic
  content occur because of media pressure. Second, using interrupted time
  series analysis, we show that media attention on communities with toxic
  content only increases the problematic behavior associated with that
  community (both within the community itself and across the platform).
  However, we find no significant difference in the impact of administrative
  interventions on subreddits with and without media pressure. Taken all
  together, this study provides evidence of a media-driven moderation strategy
  at Reddit and also suggests that such a strategy may not have a significantly
  different impact than a more proactive strategy.
 \end{abstract}

\section{Introduction}
\label{sec:introduction}

Strict platform moderation is rarely a first-order priority for newly developed
online platforms. After all, the early adopters are often homogenous with
a shared goal of nurturing the community. However, as platforms become more
mainstream and contend with a large and consistent influx of new users, each
with their own ideals and agendas, effective and timely platform moderation
becomes paramount to maintaining a civil community. Despite the absence of any
legal consequences for not effectively moderating platforms, effective
moderation is often tied to another goal of the platform -- avoiding negative
media attention so that the platform remains appealing to advertisers who
ultimately are their primary revenue source. Complicating matters, as
economically rational actors, platforms need to also account for the loss in
users and popularity as a result of platform-wide moderation decisions. This
suggests that the effectiveness of moderation on platforms might be tied to the
media's coverage of their failures as well as the costs of moderation decisions
on platform activity. The research presented in this paper investigates these
relationships on Reddit. 

\para{Reddit's history with media-driven moderation decisions.}
The story of platform moderation on Reddit appears similar to the evolutionary
trend described above. In its early days, Reddit was celebrated as the bastion
of free speech due to its minimal moderation and interference. However, as its
popularity grew over the years it found itself being criticized by outsiders
and the media for its lack of effective moderation. There have been numerous
examples of Reddit's moderation decisions being driven by media pressure
including \subreddit{The\_Donald} which was only shutdown after widespread
reporting in the media for the violent and incivil political discourse it
facilitated, \subreddit{TheFappening} which was shutdown only after reports of
its role as the facilitator in the distribution of involuntary pornography
involving celebrities, \subreddit{CoonTown} which was not banned during
Reddit's first purge of `hateful' subreddits until criticism from mainstream
media outlets \cite{Moyer-WaPo2015}, and most notably -- \subreddit{jailbait}.
The \subreddit{jailbait} subreddit was one of the earliest cases of Reddit
moderation being performed only in reaction to media attention
\cite{centivany2016values}. The subreddit featured provocative pictures of
minors and due to the lack of any rules against it, Reddit condoned its
existence even awarding it the \textit{voted best subreddit of 2008}
\cite{Chen-Jailbait2012}. In September 2011, in a segment on his show,
Anderson Cooper of CNN brought \subreddit{jailbait} to wider attention heavily
criticizing Reddit on hosting such content. Following more negative attention,
the subreddit was finally banned by administrators in October 2011. This
extremely delayed intervention led many, including the creator of the
subreddit, to speculate that the closing of the subreddit was only direct
response to the negative attention \cite{Tufekci-CITP2012}.
This speculation was further validated by the lack of administrative action
against other `bait'-type subreddits such as \subreddit{asianjailbait}. Taken
together, these anecdotes suggest that media pressure does impact Reddit's
moderation decisions. The extent of this impact is the subject of this
research.

\para{The consequences of media-driven moderation.} Aid from the media, users,
and outsiders helps platforms conduct effective moderation. By bringing
attention to egregious content and highlighting gaps in its policies, such
attention can help platforms perform difficult administrative actions and
evolve their content policy. However, over reliance on the media for moderation
may lead to several problems including: inconsistent enforcement of policies
owing to the medias own inconsistent coverage of problematic content, delayed
moderation decisions due to the fact that action is taken only after
a violation is egregious enough to warrant coverage by the media, and finally
the normalization of problematic behaviours since media coverage may only focus
on the egregious violations while ignoring the problematic behaviours leading
up to it. Our work seeks to uncover whether these consequences are also
experienced by Reddit.

\para{Our hypotheses.} This research seeks to highlight the extent to which
media reporting drives Reddit moderation decisions and the consequences it
subsequently faces. Specifically, we explore the following hypotheses.

\parait{H1: In communities with toxic content, Reddit's administrative
interventions  for violating the content policy related to toxicity occur
because of media pressure. (\Cref{sec:relationship})} We test the validity of
this hypothesis by checking if media pressure generated by a subreddit
(quantified from negative media coverage of a subreddit) mediates the
relationship between its measured levels of toxicity and administrative
interventions for violating the content policy related to toxic content. Our
analysis shows that measures of media pressure and internal pressure completely
explains any relationship between measured levels of toxicity and
administrative interventions for violating the content policy related to toxic
content. This suggests a reactionary moderation strategy.

%
\parait{H2: Prior media attention on communities which receive interventions
for toxic content: (1) increases the prevalence of problematic activity on the
platform and (2) reduces the effectiveness of the issued interventions.
(\Cref{sec:consequences})} We
now focus on subreddits which: (1) received an administrative intervention for
violating the content policy regarding toxic content and (2) received negative
media attention prior to the administrative intervention. 
For these subreddits, we conduct an interrupted time series analysis to
understand the platform-wide increase of problematic activity related to the
toxic community as a consequence of: (1) the media pressure they receive and
(2) the administrative intervention. Our analysis shows that
media pressure and interventions both increase the levels of problematic
activity. However, we find that the effects of the intervention are not
statistically different from the effects observed by the communities which
received no media pressure prior to their intervention --- \ie interventions
are not less effective when they are preceded by media attention on the
targeted community.

\section{Are Reddit's administrative interventions influenced by media
pressure?} \label{sec:relationship}
%
\para{Overview.} In this section, we explore the relationship between media
pressure and administrative interventions in the context of toxic Reddit
communities. Our focus is solely on subreddits which were banned or quarantined for violating
the content policy related to toxicity\footnote{``Rule 1: Remember the human.
Reddit is a place for creating community and belonging, not for attacking
marginalized or vulnerable groups of people. Everyone has a right to use
Reddit free of harassment, bullying, and threats of violence. Communities and
users that promote hate based on identity or vulnerability will be banned.''}.
Our hypothesis is that: \emph{(H1) In communities with toxic content, Reddit's
administrative interventions for violating the content policy related to
toxicity occur because of media pressure.} Therefore, we
seek to test whether the effect of a subreddit's measured toxicity on
administrative interventions for violating the content policy related to
toxicity occurs because of the pressure generated by the media. Put
another way: when the toxicity of two subreddits are controlled for does the
subreddit garnering more negative media attention become more likely to
receive an administrative intervention for violating the content policy related
to toxic content? If this hypothesis is valid, it would suggest that Reddit
employs a reactionary administrative strategy which delays administrative
interventions for toxic communities until media pressure forces action. 

\subsection{Methods and datasets} \label{sec:relationship:methods}

\para{Quantifying subreddit toxicity.} We quantify the toxicity of
a subreddit as the percentage of toxic content (posts and comments) present in
a subreddit. The use of this metric is supported by comments from Reddit
administrators. For example, in response to a question demanding transparency
in their administrative interventions for violations of Rule 1, \emph{u/spez}
(an administrator and co-founder of Reddit) indicated that ``high ratio'' of
hateful content was a major criteria for
interventions.\footnote{https://www.reddit.com/r/announcements/comments/hi3oht/update\_to\_our\_content\_policy/fwe83at/}
This also motivates our study of the relationship between toxicity, media
pressure, and administrative interventions for toxic content. In order to
identify toxic content, we leverage the Perspective
API\footnote{https://www.perspectiveapi.com/} --- a Google-owned tool for
identifying online toxic content. We use the Perspective API to identify
the percentage of all toxic comments and posts on a subreddit. We quantify the
toxicity of a subreddit as $T(s) = \frac{\text{\# toxic comments} \in
s + \text{\# toxic posts} \in s}{\text{\# comments} \in s + \text{\# posts} \in
s}$. We note that the Perspective API has been validated for use with Reddit
and has been leveraged to quantify subreddit toxicity in several previous
studies  \cite{mittos2020and, zannettou2020measuring} and has also been used as
a plugin to aid moderation.
\footnote{https://www.perspectiveapi.com/case-studies/}

\para{Quantifying negative media attention as media pressure.} We seek to
quantify negative attention towards subreddits from popular media outlets. We
start by identifying the number of published media articles that mention
a subreddit in a negative or critical tone. We refer to each of these articles
as a `negative media mention'. To measure the negative media mentions for
a subreddit, we use the MIT media cloud API\footnote{https://mediacloud.org/}
to obtain articles mentioning the subreddit's name. We restrict our analysis to
articles from US `top sources' and `mainstream media' sites as categorized by
the MIT media cloud. We focus the remainder of our analysis only
on articles published between 01/2015 and 04/2020. Further, for subreddits
which received an intervention we only include pre-intervention articles (\ie
those published up to the month prior to the intervention). We do this to
ensure the exclusion of articles which report the occurrence of an
intervention. Next, for each article, we use the entity-level sentiment
analysis API from the Google NLP
platform\footnote{https://cloud.google.com/natural-language} to measure the
sentiment towards the subreddit. Articles which include negative sentiments
towards the subreddit are counted as negative media mentions. We quantify the
`media pressure' towards a subreddit $s$ as $P_{media}(s)
= \frac{\text{negative media mentions of }s}{\text{total media mentions of
}s+L}$, where $L$ is the Laplace smoothing constant and is set to 10. This
metric captures the frequency of negative media mentions relative to all media
mentions received by a subreddit. 
The presence of `total media mentions' and the smoothing constant $L$
ensures that the quantified media pressure ($P_{media}$): (1) is not identical
for two subreddits $a$ and $b$, where $a$ and $b$ have similarly high ratio of
negative:total media mentions but differ significantly in their raw number of
total media mentions and (2) is not identical for two subreddits $a$ and
$b$, where $a$ and $b$ have the same number of negative media mentions but
significantly different total media mentions.

\para{Identifying subreddits receiving administrative interventions for
violating the content policy related to toxicity.} Reddit's content policy
requires communities (\ie subreddits) to adhere to eight rules
\footnote{https://www.redditinc.com/policies/content-policy}. Violation of
these rules are meant to result in administrative interventions by Reddit. In
this paper, we focus on the communities found to be in violation of \emph{Rule
1} (commonly referred to as the anti-toxicity policy). We focus on this rule
specifically because it was the subject of the most administrative
interventions during the period of this study (from 01/2015 to 04/2020).
Further, anecdotes of media-driven interventions appear to occur most
frequently for communities found to be violating this policy, perhaps due to
its subjective nature. In order to identify subreddits banned/quarantined for violations
related to the anti-toxicity content policy, we scraped the homepages of all
subreddits and identified the ones marked as banned or quarantined for
violations of the policy \footnote{Reddit provides specific violations in the
subreddit homepage when a ban occurs. See \url{www.reddit.com/r/The\_Donald} as
an example.}. In total, 120 of the 535 subreddits which received an
administrative intervention from 01/2015 to 04/2020 were targeted for the
violation of this policy. In the remainder of this paper we broadly use the
term `administrative interventions' to refer to administrative interventions
whose stated reason was a violation of the anti-toxicity content policy.

\begin{table}[]
\resizebox{\columnwidth}{!}{%
\begin{tabular}{@{}clcccc@{}}
\textbf{Dataset}    & \textbf{Label} & \textbf{Subreddits}
  & \textbf{\begin{tabular}[c]{@{}c@{}}Avg.\\ Toxicity \\($T$)\end{tabular}}
    & \textbf{\begin{tabular}[c]{@{}c@{}}Negative\\Media\\Mentions\end{tabular}}
      & \textbf{\begin{tabular}[c]{@{}c@{}}Avg. Media \\ Pressure \%age\\
      ($P_{media}$)\end{tabular}} \\
\toprule
  \multirow{2}{*}{\D{3K}} & Intervened     & 29   & \textbf{23\%} & 407
  & \textbf{18.7} \\ \cmidrule(l){2-6}
		    & Active         & 2971 & \textbf{8\% } & 644 & \textbf{0.9}  \\
        \midrule
  \multirow{2}{*}{\D{P}} & Intervened     & 120  & \textbf{24\%} & 463
  & \textbf{5.6}  \\ \cmidrule(l){2-6} 
		    & Active         & 120  & \textbf{6\%}  & 8   & \textbf{0.2}  \\
        \midrule
  \multirow{2}{*}{\D{T}} & Intervened     & 120  &         24\%  & 463
  & \textbf{5.6}  \\ \cmidrule(l){2-6} 
		    & Active         & 120  &         24\%  & 31  & \textbf{1.9}  \\
        \bottomrule
  \multirow{2}{*}{All} & Intervened     & 120  &   {\bf 24\%}  & 463
  & \textbf{5.6}  \\ \cmidrule(l){2-6} 
        & Active         & 3211  &        {\bf 9\%}  & 683  & \textbf{0.9}  \\
        \bottomrule
\\
\end{tabular}%
}
\caption{Characteristics of the datasets used in this study. Bold values
  indicate a statistically significant ($p$ < .05) difference between the
  attributes for the intervened and active groups in the corresponding
  dataset.}
\label{table:relationship:datasets}
\end{table}

\para{Datasets.}
Our data was gathered using Pushshift \cite{PushShift-Reddit} and comprised of
all the comments and posts made on Reddit during the period from 01/2015 to
04/2020. In total, this included 5B comments and 684M posts from 39M unique
users. For the analysis presented in this section, we use this data to
construct three different datasets that are described below. The
characteristics of each dataset is illustrated in
\Cref{table:relationship:datasets}.

\parait{Dataset of most active subreddits (\D{3K}):} This dataset contains all
the content (comments, posts, and media mentions) associated with the 3000 most
active subreddits between 01/2015 and 04/2020. We define activity as the
average number of monthly comments and posts made on the subreddit. For
subreddits which receive an administration intervention (referred to as
`intervened subreddits'), this average is computed only over the post-creation
and pre-intervention months that occurred within the period from 01/2015 to
04/2020. For subreddits without an administrative intervention (referred to as
`active subreddits'), this average is computed over all the post-creation
months that occurred between 01/2015 and 04/2020. In total, this dataset
contained 29 intervened subreddits and 2971 active subreddits.

\parait{Dataset of popularity-controlled subreddits (\D{P}):} This dataset
contains all the content associated with all 120 subreddits which received an 
intervention for violating the `anti-toxicity' policy between 01/2015 and
04/2020. For each of these intervened subreddits, we also include content
associated with an active subreddit that has the most similar popularity.
Popularity is measured by the average number of active users on the subreddit
each month (\ie the number of unique users making posts or comments on the
subreddit). As above, this average is only computed over the subreddit's
post-creation and pre-intervention period between 01/2015 and 04/2020.
A Kolmogorov-Smirnoff goodness-of-fit test shows that the
distributions of popularity observed within the two groups of subreddits in
this dataset (\ie active and intervened) are statistically similar ($p$ < .05).

\parait{Dataset of toxicity-controlled subreddits (\D{T}):} This dataset also
contains all the content associated with our 120 intervened subreddits.
However, the active subreddits in this dataset are obtained by matching each
intervened subreddit with the non-intervened subreddit having the most similar
toxicity ($T$) score. Similar to the previous datasets, toxicity scores were
only computed over the post-creation and pre-intervention period between
01/2015 and 04/2020. A Kolmogorov-Smirnoff goodness-of-fit test shows that the
distributions of toxicity scores observed within the two groups of subreddits
in this dataset are similar ($p$ < .05).

\subsection{Analysis and results}\label{sec:relationship:results}

\para{Overview of analyses.} 
We conduct three observational
experiments to better understand the influence of toxicity ($T$) and media
pressure ($P_{media}$) on each other and on administrative interventions for
toxic content. Each experiment builds on the previous and eventually provides
a test for \emph{H1}. 

\para{Analysis 1: What are the characteristics of intervened subreddits?} We
begin our analysis by simply comparing the distributions and means of toxicity
scores ($T$) and media pressure scores ($P_{media}$) for active and intervened
subreddits in each of our three datasets (\D{3K}, \D{P}, and \D{T}). 

\parait{Differences in distributions of $P_{media}$ and $T$:} In all three
datasets, we find that the distribution of $P_{media}$ scores is statistically
significantly different for active and intervened subreddits. Similarly, we see
statistically significant differences in $T$ scores for active and intervened
subreddits in \D{3K} and \D{P} (not in \D{T} which specifically controls for
toxicity across the two groups). Looking at the means, we see that on average
and across all three datasets, intervened subreddits have over $6\times$ higher
$P_{media}$ and $2.5\times$ higher $T$ scores than non-intervened subreddits.
Interestingly, we find that even when toxicity scores are controlled (\D{T}),
the mean $P_{media}$ score of intervened subreddits is nearly $3\times$ higher
than their equally toxic non-intervened counterparts. These results suggest
that $P_{media}$ may be more predictive (than $T$) of administrative
interventions. However, we note that only 43 of the 120
intervened subreddits had received media attention prior to their intervention.
This suggests that $P_{media}$ is not the only influence or predictor of an
intervention. A full breakdown of $T$ and $P_{media}$ scores for each group and
dataset is provided in \Cref{table:relationship:datasets}.

\parait{Predictive powers of $T$ and $P_{media}$ on administrative
interventions:} Next, we construct a logistic regression model that uses $T$
and $P_{media}$ to predict administrative interventions. We find that both
variables are statistically significant predictors of administrative
interventions with identical odds ratios of 4\% --- \ie all else equal, a unit
increase in $T$ or $P_{media}$ increases the odds of a subreddit receiving an
intervention by 4\%. This result suggests that the administrative interventions
may be influenced by both $T$ and $P_{media}$ and therefore justifies further
investigation into their relationship with each other and with administrative
interventions.

\begin{figure}[t]

\begin{subfigure}[b]{\linewidth}
         \centering
         \includegraphics[width=\linewidth]{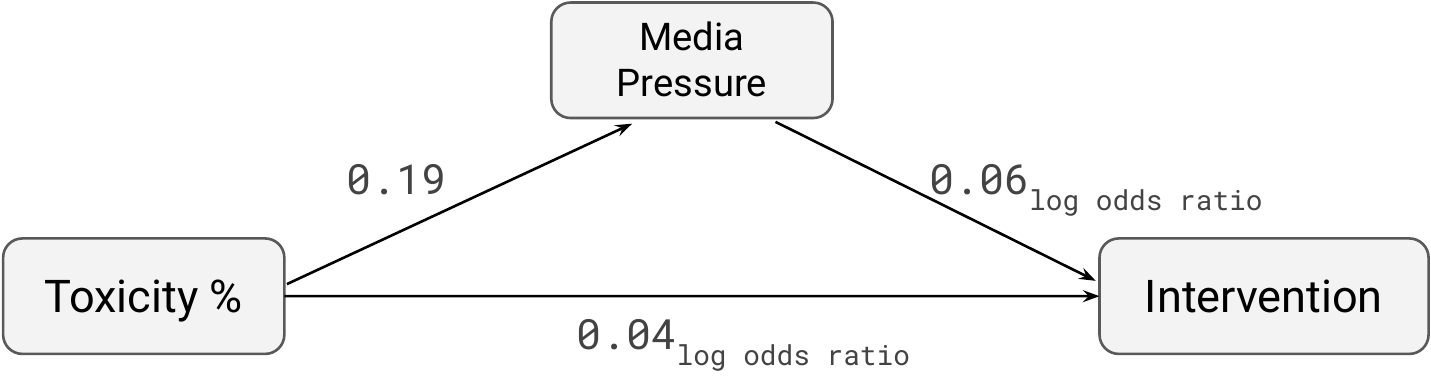}
         \caption{Mediation effects observed on \D{3K}. The direct effect ($T
         \rightarrow I$) is .04 (log odds) and the indirect effect ($T
         \rightarrow P_{media} \rightarrow I$) is .13 (log odds). Both
         effects are statistically significant ($p$ < .05).}
         \label{figure:relationship:simplemediation:D3k}
     \end{subfigure}

     \begin{subfigure}[b]{\linewidth}
         \centering
           \includegraphics[width=\linewidth]{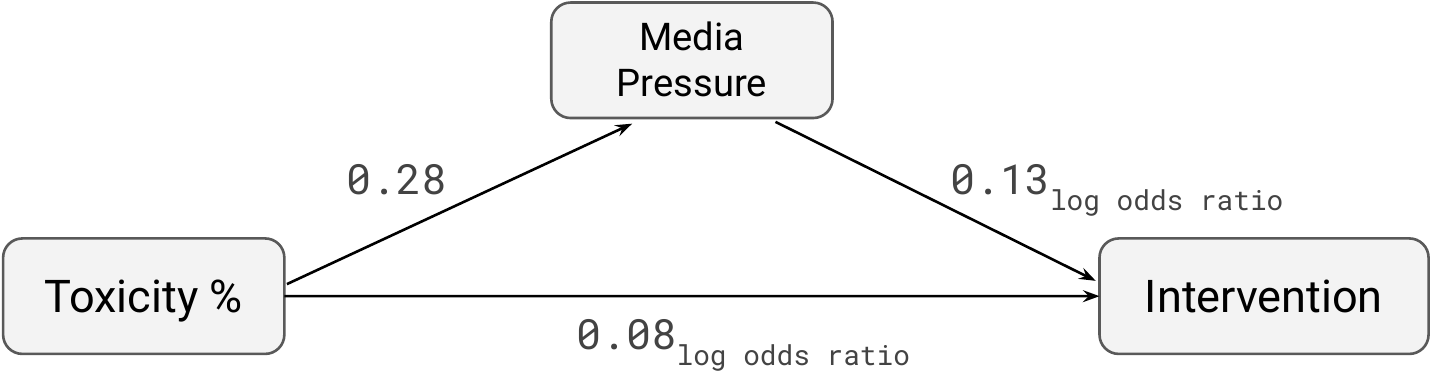}
         \caption{Mediation effects observed on \D{P}.  The direct effect ($T
         \rightarrow I$) is .08 (log odds) and the indirect effect ($T
         \rightarrow P_{media} \rightarrow I$) is .18 (log odds). Both
         effects are statistically significant ($p$ < .05).}
         \label{figure:relationship:simplemediation:Dp}
         \label{fig:H1-mediation-DP}
     \end{subfigure}
     \caption{A preliminary mediation analysis: Does $P_{media}$ mediate the
     relationship between $T$ and $I$? Solid lines indicate statistically
     significant effects. Values indicate correlation coefficients between
     variables. }
     \label{figure:relationship:simplemediation}
\end{figure}

\para{Analysis 2: Does media pressure mediate the relationship between toxicity
and administrative interventions?} Our previous results show
that $P_{media}$ and $T$ scores are predictive of administrative interventions
on subreddits.  Further, we find that $T$ is a statistically significant
predictor of $P_{media}$. Both these findings suggest the possibility of $T$
having its relationship with administrative interventions mediated by
$P_{media}$ --- \ie the effects of $T$ on administrative interventions may be
explained by $T$'s effects of $P_{media}$. We explore this with a  mediation
model.

%
\parait{Testing $P_{media}$ as a mediation variable.} We now consider
a mediation model which uses $T$ as the independent variable, an indicator
variable ($I$) to represent administrative interventions ($I_s=1$ if the
subreddit $s$ received an administrative intervention and
$I_s=0$ otherwise) as the dependent variable, and $P_{media}$ as the mediating
variable. We conduct our mediation analysis on the \D{3K} and \D{P} datasets.
Note that the \D{T} dataset cannot be used since it explicitly controls for
toxicity (the independent variable in our model) which would forcibly remove
any effects from $T\rightarrow I$.
Our models, the direct $T \rightarrow I$ effects, and indirect $T \rightarrow
P_{media} \rightarrow I$ are illustrated in
\Cref{figure:relationship:simplemediation:D3k} (for dataset \D{3K}) and 
\Cref{figure:relationship:simplemediation:Dp} (for dataset \D{p}). In both
cases, we see that the mediation occurring through $P_{media}$ is statistically
significant and that the indirect effect from $T \rightarrow P_{media}
\rightarrow I$ is substantially higher than the direct effect from $T
\rightarrow I$. In the \D{3K} dataset, a unit increase in $T$ will result in
a 4\% increase in the odds of an intervention solely due to $T$ and a 14\%
increase in the odds of an intervention because of the effect of $T$ on
$P_{media}$. Similarly, in the \D{p} dataset, a unit increase in $T$ will
result in a 8\% increase in the odds of an intervention solely due to $T$ and
a 19\% increase in the odds of an intervention because of the effect of $T$ on
$P_{media}$. Thus, we can conclude that $P_{media}$ has a partial
mediating effect on $T \rightarrow I$. 

\para{Analysis 3: Further exploring the relationship between toxicity and
administrative interventions?} We now seek to build
a complete model to explain the relationships between $T$, $P_{media}$, and
$I$. Our initial analysis which shows that $P_{media} > 0$ only in 43
communities and the existence of only a partial mediation by $P_{media}$
suggests the possibility for additional influences between $T\rightarrow I$. We
explore this possibility by incorporating several third variables into our
model: (as moderators) subreddit popularity, subreddit topic, and subreddit
profitability; and (as mediators) internal pressure and external pressure.
We use this model to analyze the \D{P} dataset since it is the most complete
and allows effects of toxicity.

\parait{Including moderating variables.} \emph{Subreddit popularity} is
a measure of the average number of active contributors to
a subreddit per month. In our model, we specifically investigate how subreddit
popularity may influence the relationship between $T \rightarrow P_{media}$ and
$T \rightarrow I$. The inclusion of popularity allows us to investigate whether
the $T \rightarrow P_{media}$ or $T \rightarrow I$ effects are significant and
stronger for subreddits of different popularity levels. 
Next, we include \emph{subreddit topic} as a moderator. For this, we leverage
TF-IDF to generate keyword vectors for each subreddit and then apply $k$-means
clustering over these vectors ($k$=8 was found to perform best for subreddits
in the \D{p} dataset). We manually label each cluster with the general
topics of subreddits contained in them. The topics identified were: sports,
politics, forums, memes, gore, porn, games, and health. All
subreddits in a cluster received the cluster's topic as its own. In our
analysis of subreddit topics, we were specifically interested in studying the
effects of subreddit topic on $T \rightarrow P_{media}$.
Finally, we introduce a \emph{subreddit profitability} variable as a moderator.
In addition to advertising revenue, Reddit is supported by Redditors' purchase
of Reddit coins
\footnote{\url{https://reddithelp.com/hc/en-us/articles/360043034252}}. These
coins allow Redditors to reward high-quality posts and comments with awards and
reactions. We estimate the amount of non-advertising revenue generated by
a subreddit by tracking the average number of awards donated to posts and
comments on the subreddit each month. This estimate is used as a proxy for
subreddit profitability. In our analysis, we are specifically interested in
understanding how subreddit profitability moderates the relationships between
$T \rightarrow I$ and $T \rightarrow P_{media} \rightarrow I$.

\parait{Introducing mediating variables.} Our complete model also seeks to
understand if the influence of pressure on administrators originating from within
the Reddit community (internal non-media pressure or $P_{int}$) and pressure on
administrators originating from other platforms (external non-media pressure or
$P_{ext}$) may mediate $T\rightarrow I$. In order to measure $P_{int}$ for
a subreddit, we obtain all pre-intervention and post-creation comments made on
Reddit between 01/2015 and 04/2020 which mention the specific subreddit in
a negative sentiment. Then we set $P_{int} = \frac{\text{negative comment
mentions of }s}{\text{total comment mentions of }s+L}$. We quantify external
pressure for a subreddit by gathering all pre-intervention and post-creation
tweets made on Twitter between 01/2015 and 04/2020 which mention the specific
subreddit in a negative sentiment. Same as before, we set $P_{ext}
= \frac{\text{negative Twitter mentions of }s}{\text{total Twitter mentions of
}s+L}$. We select Twitter as our proxy for external pressure due to its
ubiquity, size, and prominence in the activist community. 

\begin{figure}[t]
    \centering
    \includegraphics[width=\linewidth]{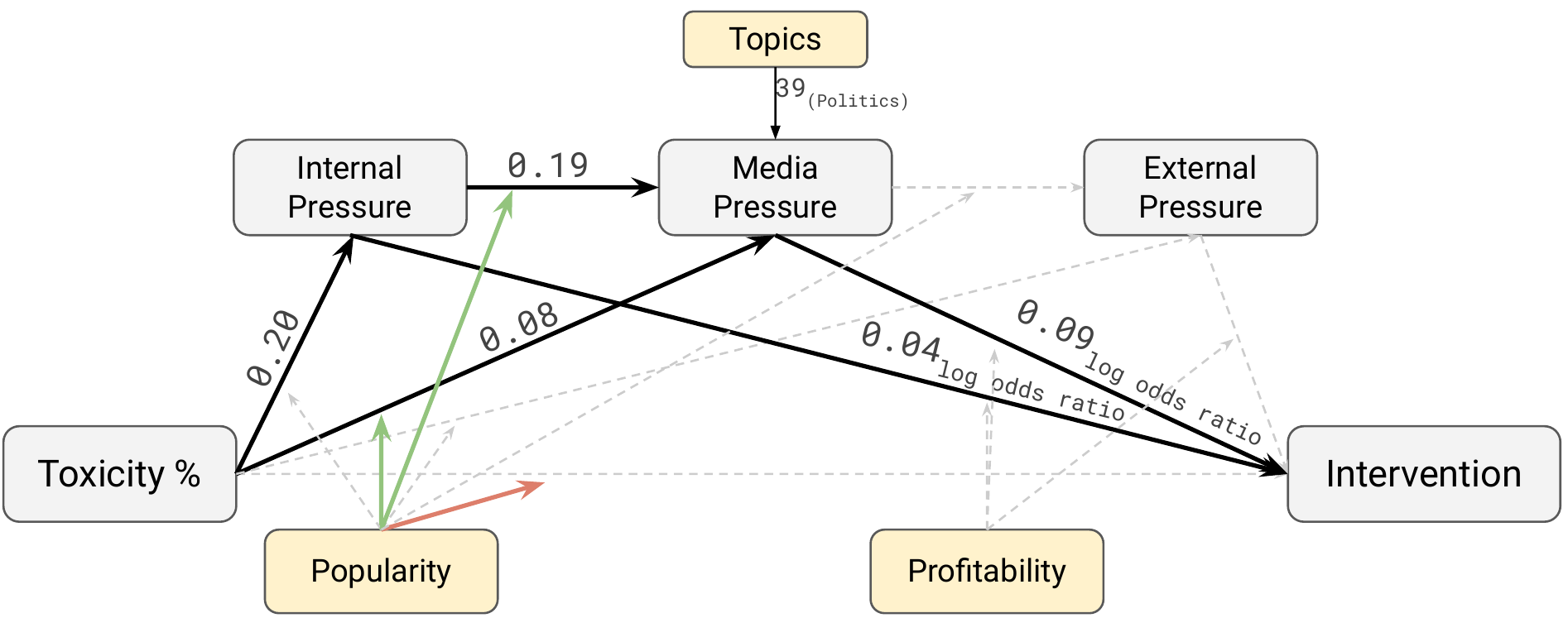}
    \caption{A complete mediation analysis. Solid
    lines indicate statistically significant effects and dashed lines indicate
    insignificant effects. Values indicate the correlation coefficients between
    variables. Variables in yellow boxes were included as moderators. Green
    and red arrows indicate a statistically significant amplifying and
    dampening moderation effect, respectively.}
    \label{figure:relationship:e3mediation}
\end{figure}

\parait{Pathways to administrative interventions.} The results of our complete
mediation analysis are illustrated in \Cref{figure:relationship:e3mediation}. 
First, we see that \emph{$P_{int}$ and $P_{media}$ \underline{completely mediate} the
    relationship between $T$ and $I$.} The inclusion of $P_{int}$ and
    $P_{media}$ as mediators between $T$ and $I$ cause the direct effect $T
    \rightarrow I$ to become insignificant. This allows us to conclude that any
    effect that $T$ has on $I$ is only because of its effect on $P_{int}$ and
    $P_{media}$. Analyzing the pathways to influence $I$, we see that all the
    indirect effects through $P_{int}$ and $P_{media}$ are statistically
    significant. Of these paths, the indirect
    effect from $T \rightarrow P_{int} \rightarrow P_{media} \rightarrow I$ is
    found to be the strongest with a unit increase in $T$ resulting in a 2.3\%
    increase in the odds of an intervention through this path. Smaller effects
    are observed on the $T \rightarrow P_{media} \rightarrow I$ and $T
    \rightarrow P_{int} \rightarrow I$ paths where a unit increase in $T$
    increases the odds of intervention by 1.2\% and 1.7\%, respectively.
Our model also shows that \emph{subreddit popularity moderates relationships with $P_{media}$ and
    the effect of $T$ on $I$.} Specifically, we find that subreddit popularity
    is a statistically
    significant amplifier in the $T \rightarrow P_{media}$ and $P_{int}
    \rightarrow P_{media}$ relationships --- \ie the influence of $T$ and
    $P_{int}$ on $P_{media}$ is higher for more popular subreddits than less
    popular ones when toxicity or $P_{int}$ are controlled for. This
    intuitively makes sense --- after all, media outlets' interest in covering
    a subreddit is likely related to the popularity of the subreddit. We
    also find that the effect of $T \rightarrow I$ reduces as popularity
    increases and this effect, although small, becomes statistically
    significant for subreddits with popularity in the 84th percentile and
    higher. This finding suggests a marginal hesitation to apply
    administrative interventions to more popular subreddits when toxicity is
    controlled for.
We note that \emph{Subreddit topic, subreddit
    profitability, and external pressure yielded no statistically significant
    influences in our model.} This suggests that subreddit topic does not
    influence     media pressure when toxicity is controlled, subreddit
    profitability never     influences administrative interventions (through
    the direct or indirect     path), external pressure is not influenced by
    media pressure or toxicity,     and external pressure does not influence
    administrator intervention     decisions.

\para{Takeaways.} Our results confirm our original hypothesis that in
communities with toxic content ($T$), Reddit's administrative interventions
for violating the content policy related to toxicity occur ($I$) because of
media pressure ($P_{media}$). However, our analysis shows that the mediating
effect of media pressure ($T \rightarrow P_{media} \rightarrow I$) does not
completely explain the relationship between $T$ and $I$. We find that
incorporating the effects of internal pressure ($P_{int}$) in our model yields
two additional statistically significant pathways: $T \rightarrow P_{int}
\rightarrow I$ and $T \rightarrow P_{int} \rightarrow P_{media} \rightarrow I$
whose addition completely explains any effect from $T$ to $I$. Taken all
together, this suggests a reactionary moderation strategy in which any
administrative interventions handed out for toxic content are driven by
internal pressure from Redditors and media pressure from negative media
attention.

\section{What are the consequences of a media-driven intervention strategy?}
\label{sec:consequences}

\para{Overview.}
Thus far, our analysis has demonstrated that administrative interventions
for toxic content are largely driven by internal and media pressure. We now
seek to understand whether such reactionary administrative
intervention strategies are effective at curbing problematic activities (that
are associated with the target subreddit) across the platform. Our hypothesis
is that: \emph{(H2) Prior media attention on communities which receive
interventions for toxic content: (1) increases the prevalence of problematic
activity on the platform and (2) reduces the effectiveness of the issued
interventions.} We test this hypothesis using an interrupted time series
analysis to check whether community-specific increases in user growth rates
and platform-wide increases in problematic discourse (that is associated with
the intervened subreddit) occur as a consequence of the media pressure they
receive and the administrative intervention they are handed out. If
part (1) of this hypothesis is valid, it suggests that media attention
on a problematic community increases the prevalence of the problematic
discourse within the community and across the platform. If part (2) of this
hypothesis is valid, it suggests that media-driven interventions are less
effective at curbing the spread of problematic discourse across the platform
than their non-(media)impacted counterparts.

\subsection{Methods} \label{sec:consequences:methods}

\para{Tracking growth rates within an intervened subreddit.} For each of the
120 intervened subreddits in our dataset, we compute the monthly `growth' of
the community. 
This growth for a given month is computed by counting the number of unique
users that made their first contributions to the community during that month.
Put another way, this measures the number of new contributors to a community
each month. This metric is used to identify the impact that media coverage has
on intervened communities. Note that this metric cannot be used to identify the
impact of administrative interventions since the community itself becomes
inactive after the intervention.
We hypothesize that media attention results in an
increase in the growth rate of a community --- \ie more users begin directly
participating in the problematic discourse as a result of the media attention.

\para{Identifying and tracking problematic discourse of an intervened
community.} For our analysis, we seek to measure if the ``problematic
discourse'' of an intervened community begins to spread across the platform as 
a consequence of media attention and administrative interventions on the
community. This requires us to identify and track this problematic discourse.
We do this by using the vocabulary unique to the intervened subreddit as
a proxy for the problematic discourse occurring on it. By tracking the
prevalence of this unique vocabulary on other subreddits, we effectively
measure the adoption of the problematic vocabulary across the platform. We
derive the unique vocabulary associated with a community using the
\emph{Sparse Additive Generative Model (SAGE)} \cite{SAGE}. SAGE extracts
keywords that are unique to our intervened subreddit relative to a set
of reference subreddits (the default subreddits in our case). Using this
process, we extract the 500 most unique keywords for each intervened community
and manually confirming their relevance and specificity to the intervened
community. For example, (\textit{fakecel, truecel, zyrros, femoid...}) were the
extracted from \subreddit{Incels} and
(\textit{eyethespy, thankq, ibor...}) were extracted from
\subreddit{greatawakening}. 
We then count the frequency of occurrence of these keywords across
the remainder of the platform (\ie excluding the intervened subreddit itself)
for each month. 
We note that this approach has also been used in prior work measuring 
the spread of ideologies \cite{eshwar2017you}. 

\para{Identifying the effects of media coverage and administrative
interventions using interrupted time series analysis.} Interrupted time series
analysis test for any significant changes in the rate of a given variable after
an event of interest occurs. It models the time series prior to the event and
forecasts the time series after the event. If there is a statistically
significant difference in the forecasted and actual time series after the
event, the event is said to have an effect on the variable being tracked. We
run three interrupted time series analyses for each of our 120 intervened
communities: (1) for communities experiencing pre-intervention media attention,
using community growth rates as the variable and media attention as the event,
(2) for communities experiencing pre-intervention media attention, using the
growth in prevalence of intervened subreddit vocabulary on other subreddits as
the variable and media attention as the event, and (3) for all intervened
subreddits, using the prevalence of intervened subreddit vocabulary on other
subreddits as the variable and the administrative intervention as the event.
All together, these analyses will identify the impact of media attention on
problematic discourse and activity.

\para{Comparing the effectiveness of media-driven interventions with
interventions not impacted by media coverage.} Finally, we split our dataset of
120 intervened subreddits into those which generated media pressure (treatment)
and those that did not (control). We then compare the effects of administrative
interventions on these two groups with a focus on the percentage increase in
occurrences of their vocabulary on other subreddits. A statistically
significant difference in this variable between the two groups would suggest
the possibility that media-driven interventions are less effective at curbing
the spread of problematic discourse and ideologies.

\subsection{Analysis and results}

\para{Overview of analyses.} To test our hypothesis ($H2$), we conduct three
different analyses with each testing the impact of media pressure and
interventions on subreddit growth and spread of problematic discourse.

%
%
%

\para{Analysis 1: For toxic communities, what is the impact of negative media
attention on subreddit growth?} We now focus on the subset of our intervened
subreddits which received negative media attention prior to their
administrative intervention for toxic content --- 43 in total. We conduct an
interrupted time series analysis to test whether there was anomalous community
growth (quantified by the rate of new creators joining the community) after the
first time they received media attention. Across all 43 intervened  subreddits
with prior media attention, we see that 28 had statistically significant
increases in user growth after the first time they received media attention.
The average growth observed was 439\%. Despite these alarming increases, the
impact of media attention appears disparate for different communities --- \eg
\subreddit{greatawakening} grew 4174\% while \subreddit{Mr\_Trump} only grew
42\% (both are statistically significant from our interrupted time series
analysis). Grouping the subreddits by their topics, we see patterns emerge ---
groups that received the most negative media  attention (subreddits in the
manosphere, qanon, and extremist ideology categories) also had the largest
growth rates from negative media attention. A subset of these results, grouped
by `subreddit topic' are reported in \Cref{table:consequences:results} in the
\emph{User growth (post-media)} column. Our findings provide
evidence that negative media attention increases the growth rate for toxic
communities.

\para{Analysis 2: For toxic communities, what is the impact of negative media
attention on the spread of problematic community vocabulary?} Once again we 
focus on the 43 intervened subreddits which received negative media attention
prior to their interventions. We use an interrupted time series analysis to
test whether the vocabulary of problematic subreddits is more commonly adopted
across the platform after the first time they received media attention. The
interrupted time series analysis returns statistically significant results if,
given prior data, the growth of usage of the vocabulary on other subreddits is
anomalous after the first media attention. We find that 20 of our 43 subreddits
recorded statistically significant changes in the adoption of their
vocabulary across the platform. The average increase across all 43 subreddits
was 128\%. Once again, we find that the effects are disparate across
communities -- \eg \subreddit{shortcels} experienced an increase of 1270\%
while \subreddit{Mr\_Trump} experienced a decrease of 79\%. Specifically
analyzing subreddits by their category, we find that only subreddits in the
`manosphere' experienced a consistent and significant increase in their
vocabulary adoption rates after the first time they received media attention.
A subset of these results are reported in the \emph{Voc. growth (post-media)}
column in \Cref{table:consequences:results}. Our findings show that
media attention results in increased adoption of an toxic
community's vocabulary across the platform.

\para{Analysis 3: What is the impact of administrative interventions on the
spread of problematic community vocabulary?} We use an interrupted time series
analysis to test whether the growth in usage vocabulary of a problematic
subreddit across the platform varies depending on whether the subreddit
received media attention or not. On average, across all 120 intervened
subreddits we find that 52 subreddits had a statistically significant change in
vocabulary adoption across the platform. Of these, 26 had received media
attention prior to the intervention and 26 had not. The average increase
observed in the subreddits that received media attention was 321\% and
348\% for those that did not. We note that the difference between the two
groups was not found to be statistically significant. Breaking down our results
by subreddit topic, we find that subreddits in the manosphere were once again
found to have their vocabulary consistently adopted across Reddit even after
the intervention. This breakdown is illustrated in the \emph{Voc. growth
(post-int.)} column in \Cref{table:consequences:results}. Our findings show
that subreddits which receive interventions content see their
vocabulary being adopted across the platform after an intervention, regardless
of whether they received prior media attention or not.


\para{Takeaways.} We validated one of our hypotheses
($H2(1)$) that media attention on problematic communities increases the
user growth rate in the community itself and increases the adoption of the
community's vocabulary across the platform. Our findings were unable to
validate our second hypothesis ($H2(2)$) that prior media attention on
problematic communities reduced the effectiveness of administrative
interventions. All together, our study allows us to conclude that media
attention on a problematic community does lead to an increase in problematic
activity within and outside the community itself. However, reactionary
administrative interventions do not appear to have a significantly different
impact on the communities which receive media attention.

\begin{table}[]
\resizebox{\columnwidth}{!}{%
\begin{tabular}{cllll}
\toprule
\multicolumn{1}{l}{\textbf{Topic}} &
  \textbf{Communities} &
  \textbf{\begin{tabular}[c]{@{}l@{}}User growth\\(post-media)\end{tabular}} &
    \textbf{\begin{tabular}[c]{@{}l@{}}Voc. growth\\(post-media)\end{tabular}} &
      \textbf{\begin{tabular}[c]{@{}l@{}}Voc. growth\\(post-int.)\end{tabular}}  \\ \hline \\ [-1.5ex] 
\multirow{8}{*}{\rotatebox[origin=c]{90}{\textbf{Manosphere}}}
				     & \text{Incels}      & \textbf{+512\%}  & \textbf{+216\%}  & \textbf{+205\%}  \\
                                     & \text{Braincels}             & \textbf{+1647\%} & \textbf{+231\%}  & \textbf{+228\%}  \\
                                     & \text{shortcels}             & -100\%           & \textbf{+1270\%} & \textbf{+1259\%} \\
                                     & \text{TheRedPill}            & \textbf{+32\%}   & \textbf{+219\%}  & \textbf{+250\%}  \\
                                     & \text{MGTOW}                 & \textbf{+212\%}  & \textbf{+213\%}  & \textbf{+473\%}  \\
                                     & \text{JustBeWhite}           & \textbf{-}       & \textbf{-}       & \textbf{+217\%}  \\
                                     & \text{milliondollarextreme}  & \textbf{-}       & \textbf{+283\%}  & \textbf{+283\%}  \\
                                     & \text{CringeAnarchy}         & \textbf{+124\%}  & \textbf{+94\%}   & \textbf{+131\%}  \\ [-1.5ex] \\ \hline \\ [-1.5ex]
\multirow{8}{*}{\rotatebox[origin=c]{90}{\textbf{QAnon}}}
				     & \text{greatawakening} & \textbf{+4174\%} & \textbf{-29\%}   & +46\%            \\
                                     & \text{TheCalmBeforeTheStorm} & \textbf{-}       & -                & -2\%             \\
                                     & \text{uncensorednews}        & \textbf{+285\%}  & \textbf{-59\%}   & \textbf{-7\%}    \\
                                     & \text{TheNewRight}           & \textbf{+56\%}   & +11\%            & +43\%            \\
				     & \text{The\_Donald}           & \textbf{+524\%}  & \textbf{+124\%}  & +46\%            \\
                                     & \text{Right\_Wing\_Politics} & \textbf{+401\%}  & +44\%            & +73\%            \\
                                     & \text{new\_right}            & \textbf{+401\%}  & -92\%            & +23\%            \\
                                     & \text{Mr\_Trump}             & \textbf{+42\%}   & \textbf{-79\%}   & \textbf{+284\%}  \\ [-1.5ex] \\ \hline \\ [-1.5ex]
\multirow{25}{*}{\rotatebox[origin=c]{90}{\textbf{Extremist groups}}}
				     & \text{The\_Donald} & \textbf{+524\%}  & \textbf{+124\%}  & +46\%            \\
                                     & \text{TheNewRight}           & \textbf{+56\%}   & +11\%            & +43\%            \\
                                     & \text{DebateAltRight}        & -                & -                & +125\%           \\
                                     & \text{WhiteRights}           & \textbf{+40\%}   & -45\%            & \textbf{+32\%}   \\
                                     & \text{Mr\_Trump}             & \textbf{+42\%}   & \textbf{-79\%}   & \textbf{+284\%}  \\
                                     & \text{Physical\_Removal}     & \textbf{+412\%}  & -32\%            & -68\%            \\
                                     & \text{Right\_Wing\_Politics} & \textbf{+401\%}  & +44\%            & +73\%            \\
                                     & \text{RightwingLGBT}         & \textbf{+562\%}  & +22\%            & +102\%           \\
				     & \text{european}              & -                & -                & \textbf{+312\%}  \\
                                     & \text{SargonofAkkad}         & -                & -                & -6\%             \\
                                     & \text{SubforWhitePeopleOnly} & -                & -                & +53\%            \\
                                     & \text{TheCalmBeforeTheStorm} & -                & -                & -2\%             \\
                                     & \text{ImGoingToHellForThis}  & -                & -                & \textbf{+577\%}  \\
				     & \text{CringeAnarchy}         & \textbf{+124\%}  & \textbf{+94\%}   & \textbf{+131\%}  \\
                                     & \text{The\_Europe}           & +15\%            & -41\%            & +45\%            \\
				     & \text{uncensorednews}        & \textbf{+285\%}  & \textbf{-59\%}   & \textbf{-7\%}    \\
                                     & \text{altright}              & -                & -                & -29\%            \\
                                     & \text{antifa}                & -                & -                & +182\%           \\
                                     & \text{europeannationalism}   & \textbf{+130\%}  & \textbf{-41\%}   & \textbf{+15\%}   \\
				     & \text{greatawakening}        & \textbf{+4174\%} & \textbf{-29\%}   & +46\%            \\
                                     & \text{milliondollarextreme}  & \textbf{-}       & \textbf{+283\%}  & \textbf{+283\%}  \\
                                     & \text{new\_right}            & \textbf{+401\%}  & -92\%            & +23\%            \\
                                     & \text{smuggies}              & -                & -                & \textbf{+25\%}   \\
                                     & \text{toosoon}               & -                & -                & +115\%           \\
                                     & \text{ChapoTrapHouse}        & \textbf{+563\%}  & \textbf{+177\%}  & \textbf{+3\%}    \\
                                     & \text{whitebeauty}           & -51\%            & +111\%           & \textbf{+137\%}  \\ [-1.5ex] \\ \hline \\ [-1.5ex]
\multirow{1}{*}{\rotatebox[origin=c]{90}{\textbf{}}}
				     & \textbf{Average}    & +439\% (28*) & +128\%  (20*)             & +321\%  (52*)  \\
\bottomrule
\end{tabular}%
}
\caption{\label{tab:h2-results} (Partial) Results for impact of media attention
and interventions on user growth and vocabulary adoption rate. \textbf{Bold}
values denote a statistically significant
difference in the forecasted and actual time series (p < 0.05).
Subreddits are grouped by their manually assigned category. Values in brackets
in the {\bf Average} row denote the number of statistically significant
changes.} 
\label{table:consequences:results}
\end{table}

\section{Related work} \label{sec:related}
Our research was influenced by and makes contributions to research that can
broadly be classified into two categories: platform moderation strategies and
their consequences; and the influence of externalities on platform moderation.

\para{Platform moderation strategies and their consequences.} 
The dilemma of how to moderate effectively without resorting to extreme
restrictions on discourse is not new to platforms as they increasingly find
themselves grappling with challenges arising from being too strict or too
lenient. Angwin \cite{stealing-myspace} highlighted how restrictions and
moderation on Friendster led to mass user migrations to more lenient platforms
such as MySpace. Conversely, overly lenient moderation also presents problems
for platforms. For example, the failure to address trolls and misogynistic
content led to a loss of users along with the withdrawal of several offers to
purchase and invest in Twitter \cite{disney-salesforce}. Increasingly, however,
we find platforms offering moderation strategies as a commodity: some advertise
increased safety and protection for its users (\eg Reddit and Twitter) while
others advertise no restrictions on discourse (\eg Gab, Parler, and 4chan).
Several studies, detailed below, have shown the former to suffer from
inconsistency in moderation while the complete lack of moderation in the latter
has been found to encourage extremism and toxicity \cite{zannettou2018gab,
hine2017kek}.

%
\parait{The challenge of consistent and timely moderation.}
Numerous works have tracked discourse on platforms, specifically to measure the
effectiveness of community-level interventions to suppress dangerous discourse.
Early research conducted on Reddit \cite{eshwar2017you} showed the
effectiveness of interventions applied to \subreddit{fatpeoplehate} and
\subreddit{coontown}. The study revealed a significant downturn in the amount
of incivility amongst community members after the intervention was applied.
However, this finding has been contradicted by several recent studies which
have shown users and discourse from banned communities migrating to newer
communities while maintaining or increasing their incivility
\cite{habib2019act, ali2021understanding}. Habib \etal hypothesize that
increasing inconsistency by platform administrators may be the reason for this
contradiction. Our research which suggests a reactionary moderation strategy
supports this hypothesis.
Researchers have highlighted that inconsistencies associated with
moderation may be attributed to the high cost and inherently poor scalability
of human moderation and have proposed machine-learning based tools to assist
moderators (Reddit moderators, specifically) identify communities at risk of
violating platform rules \cite{habib2019act, 2019crossmod}. Additionally,
relying on human moderators as a first line of defense has been shown to have
a severe effect of their mental health \cite{We-are-the-nerds,
roberts2014behind, wohn2019volunteer}. 

\parait{The consequences of inconsistent moderation.}
The effects of moderation inconsistencies have been found to be
substantial. In the context of the 2016 US Presidential elections, several
researchers \cite{benkler2018network, allcott2017social} found that discourse
on social media platforms played a significant role in amplifying propaganda
and fake news. These problems continue to arise today as online platforms
provide a home for fringe elements promoting violent or problematic conspiracy
theories. Failure to act effectively against such harmful ideologies by way of
timely moderator interventions has been shown to result in the development of
more extreme ideologies amongst community members. For example, researchers
\cite{mamie2021anti} showed that anti-feminist communities acted as a pathway
to more radical alt-right communities. Further, the recent attack on the US
Capitol and protests in Charlottesville that resulted in multiple deaths are
both known to have been planned in online communities including large platforms
such as Twitter and Parler \cite{prabhu2021capitol}. The importance of timely
interventions on toxic content has been further highlighted by Scrivens \etal
\cite{scrivens2020examining} who showed that there existed a gradual increase
in the approval of toxic content in response to consistent toxic posting by
community members. These results are in line with other studies showing how
communities can become more extreme over time \cite{simi2015american,
wojcieszak2010don, caiani2015transnationalization, wright2020pussy,
ribeiro2020evolution}.

%
%


\para{External forces influencing platform moderation.}
Platform moderation does not operate without influence from external
(particularly, economic and regulatory) forces. Numerous research efforts have
analyzed the impact of the online advertising ecosystem on platform moderation.
Bozarth \etal \cite{bozarth2021market} show how many fake news websites are
mostly funded by top-tier advertising firms and an effective strategy towards
combating fake news would be to have these advertisers blacklist these sites.
Braun \etal \cite{braun2019activism} showed how the `Sleeping Giants', an
activist group, strategically reported events of misinformation and racism to
brands and advertisers (rather than the platforms themselves) in an effort to
pressure them to withdraw their advertisements. This direct impact on the
revenue streams of online platforms was found to cause changes in the
moderation of misinformation and racist content. Along a similar vein, in 2019,
YouTube experienced
a series of boycotts from advertising agencies and brands in retaliation to the
proliferation of toxic content. This event, now known as the `Adpocalopyse'
resulted in a large number of changes in YouTube's content policies, comment
moderation, as well as video monetization policies \cite{kumar2019algorithmic,
caplan2020youtube}. These studies reflect the impact that
pressure from advertisers can have on the moderation policies of online
platform. Our work suggests that Reddit may not be an exception. 

\section{Discussion and conclusions} \label{sec:discussion}

\para{Limitations and challenges.} This work is fundamentally a best-effort
study to understand one aspect of the relationship between platform
administrators and the media, using observational data. Consequently, each of
our contributions has their own limitations. First, given the use of
observational data and our inability to experimentally manipulate media
pressure, we are unable to make strong causal claims regarding the relationship
between media pressure and administrative interventions. This resulted in our
need to frame a weaker hypothesis. We note, however, despite much debate
regarding the use of mediation analyses for making causal inferences, the
approach has been leveraged for precisely this purpose in many prior studies
and one could argue that our models satisfy all the criteria required to make
a causal inference \cite{pearl2014interpretation, Pieters-JCR2017}. Next, our
study required us to develop proxies for several parameters such as subreddit
profitability, topics, and external pressure. It is unclear if our analysis
found no impact from these variables due to the inaccuracy of our proxies or
the actual absence of effects from them. We note that in the absence of
ground-truth, however, one can only make best-effort approximations. Finally,
our study is also limited by our decision to treat the
media attention and interventions applied on each subreddit as independent
events. This might have implications in scenarios where one subreddit receives
negative media attention and this results in the closure of multiple related
communities (\eg Reddit banned five communities  associated with encouraging
self-harm on the same day). However, the alternate decision (grouping all
subreddits receiving an intervention together as a single class) is also
fraught with challenges that arise from the assumption that all
simultaneous interventions occur due to the same effect.

\para{Takeaways and implications.} At a high-level, our study provides evidence
of: (1) a reactionary (media- and internal-) pressure-driven administrative
strategy being leveraged by Reddit, (2) the harms of giving media attention to
toxic communities, and (3) the statistically similar (in)effectiveness of
media- and non-media driven administrative interventions. Each of these
findings has profound implications for platform administrators and media
outlets. 

\parait{Implications for platforms.} As online social platforms increasingly
find their communities becoming the originators and propagators of toxic and
harmful content, calls to regulate them have started emerging. Particularly
relevant is \S230 of the US Communications and Decency Act which grants
complete immunity to online platforms for publishing or censoring speech on
their platforms --- \ie \S230 guarantees no judicial consequences for
moderation and administration decisions. Changes to this regulation have been
proposed by both sides of the American political spectrum and, if enacted, are
expected to have severe implications for moderation strategies employed by
platforms such as Reddit. For example, any change which results in liability
for publishing a users' toxic content will likely render reactionary
administrative strategies, such as the one uncovered in our work, untenable.
Further, although our findings suggest no significant difference in the
effectiveness of interventions driven by media attention and otherwise, they do
provide evidence that reactionary interventions do facilitate an increase in
problematic behavior across the platform. Both these findings suggest the
benefits of investing in and adopting proactive intervention strategies.

\parait{Implications for media outlets.} Our study simultaneously highlights
the importance of and the dilemma faced by the media in platform moderation. On
the one hand, in the presence of reactionary platform administration and the
absence of regulatory demands, it is imperative that the media hold platforms
accountable for their administrative decisions. On the other hand, our findings
also show that shining the media spotlight on problematic communities 
results in the growth and spread of the problematic activity. Thus, it remains
unclear how media outlets should proceed --- must they continue to hold
platforms accountable or should they avoid publicizing problematic communities?
Journalists have faced similar dilemmas in the past while negotiating reporting
on hate crimes, suicides, and school shootings where they are faced with the
consequences of possibly inspiring ``copycat'' behavior. In each such case,
institutions of journalism such as the Society for Professional Journalists,
the Poynter Institute, Thomson Reuters, and others have sought input from
a variety of stakeholders in order to develop guidelines or ``best practices''
for these reports. Our research suggests the need for and value of such
guidelines for reporting toxic online content and communities.

\balance

\bibliographystyle{aaai21}
\bibliography{moderation}

\begin{thebibliography}{35}
\providecommand{\natexlab}[1]{#1}
\providecommand{\url}[1]{\texttt{#1}}
\providecommand{\urlprefix}{URL }
\expandafter\ifx\csname urlstyle\endcsname\relax
  \providecommand{\doi}[1]{doi:\discretionary{}{}{}#1}\else
  \providecommand{\doi}{doi:\discretionary{}{}{}\begingroup
  \urlstyle{rm}\Url}\fi

\bibitem[{Ali et~al.(2021)Ali, Saeed, Aldreabi, Blackburn, De~Cristofaro,
  Zannettou, and Stringhini}]{ali2021understanding}
Ali, S.; Saeed, M.~H.; Aldreabi, E.; Blackburn, J.; De~Cristofaro, E.;
  Zannettou, S.; and Stringhini, G. 2021.
\newblock Understanding the Effect of Deplatforming on Social Networks.
\newblock In \emph{13th ACM Web Science Conference 2021}.

\bibitem[{Allcott and Gentzkow(2017)}]{allcott2017social}
Allcott, H.; and Gentzkow, M. 2017.
\newblock Social media and fake news in the 2016 election.
\newblock \emph{Journal of economic perspectives} .

\bibitem[{Angwin(2009)}]{stealing-myspace}
Angwin, J. 2009.
\newblock \emph{Stealing MySpace: The battle to control the most popular
  website in America}.
\newblock Random House.

\bibitem[{Baumgartner et~al.(2020)Baumgartner, Zannettou, Keegan, Squire, and
  Blackburn}]{PushShift-Reddit}
Baumgartner, J.; Zannettou, S.; Keegan, B.; Squire, M.; and Blackburn, J. 2020.
\newblock The pushshift reddit dataset.
\newblock In \emph{Proceedings of the International AAAI Conference on Web and
  Social Media}.

\bibitem[{Benkler, Faris, and Roberts(2018)}]{benkler2018network}
Benkler, Y.; Faris, R.; and Roberts, H. 2018.
\newblock \emph{Network propaganda: Manipulation, disinformation, and
  radicalization in American politics}.
\newblock Oxford University Press.

\bibitem[{Bozarth and Budak(2021)}]{bozarth2021market}
Bozarth, L.; and Budak, C. 2021.
\newblock Market forces: Quantifying the role of top credible ad servers in the
  fake news ecosystem.
\newblock In \emph{Proceedings of the International AAAI Conference on Web and
  Social Media}, volume~15, 83--94.

\bibitem[{Braun, Coakley, and West(2019)}]{braun2019activism}
Braun, J.~A.; Coakley, J.~D.; and West, E. 2019.
\newblock Activism, advertising, and far-right media: The case of sleeping
  giants.
\newblock \emph{Media and Communication} 7(4).

\bibitem[{Caiani and Kr{\"o}ll(2015)}]{caiani2015transnationalization}
Caiani, M.; and Kr{\"o}ll, P. 2015.
\newblock The transnationalization of the extreme right and the use of the
  Internet.
\newblock \emph{International Journal of Comparative and Applied Criminal
  Justice} .

\bibitem[{Caplan and Gillespie(2020)}]{caplan2020youtube}
Caplan, R.; and Gillespie, T. 2020.
\newblock Tiered governance and demonetization: The shifting terms of labor and
  compensation in the platform economy.
\newblock \emph{Social Media+ Society} .

\bibitem[{Centivany(2016)}]{centivany2016values}
Centivany, A. 2016.
\newblock Values, ethics and participatory policymaking in online communities.
\newblock \emph{Proceedings of the Association for Information Science and
  Technology} .

\bibitem[{Chandrasekharan et~al.(2019)Chandrasekharan, Gandhi, Mustelier, and
  Gilbert}]{2019crossmod}
Chandrasekharan, E.; Gandhi, C.; Mustelier, M.~W.; and Gilbert, E. 2019.
\newblock Crossmod: A cross-community learning-based system to assist reddit
  moderators.
\newblock \emph{Proceedings of the ACM on human-computer interaction} (CSCW).

\bibitem[{Chandrasekharan et~al.(2017)Chandrasekharan, Pavalanathan,
  Srinivasan, Glynn, Eisenstein, and Gilbert}]{eshwar2017you}
Chandrasekharan, E.; Pavalanathan, U.; Srinivasan, A.; Glynn, A.; Eisenstein,
  J.; and Gilbert, E. 2017.
\newblock You can't stay here: The efficacy of reddit's 2015 ban examined
  through hate speech (CSCW).

\bibitem[{Chen(2012)}]{Chen-Jailbait2012}
Chen, A. 2012.
\newblock Unmasking Reddit's Violentacrez, The Biggest Troll on the Web.
\newblock Gawker.

\bibitem[{Eisenstein, Ahmed, and Xing(2011)}]{SAGE}
Eisenstein, J.; Ahmed, A.; and Xing, E.~P. 2011.
\newblock Sparse additive generative models of text.
\newblock In \emph{Proceedings of the 28th international conference on machine
  learning (ICML-11)}. Citeseer.

\bibitem[{Habib et~al.(2019)Habib, Musa, Zaffar, and Nithyanand}]{habib2019act}
Habib, H.; Musa, M.~B.; Zaffar, F.; and Nithyanand, R. 2019.
\newblock To Act or React: Investigating Proactive Strategies For Online
  Community Moderation.
\newblock \emph{arXiv preprint arXiv:1906.11932} .

\bibitem[{Hine et~al.(2017)Hine, Onaolapo, De~Cristofaro, Kourtellis,
  Leontiadis, Samaras, Stringhini, and Blackburn}]{hine2017kek}
Hine, G.; Onaolapo, J.; De~Cristofaro, E.; Kourtellis, N.; Leontiadis, I.;
  Samaras, R.; Stringhini, G.; and Blackburn, J. 2017.
\newblock Kek, cucks, and god emperor trump: A measurement study of 4chan’s
  politically incorrect forum and its effects on the web.
\newblock In \emph{Proceedings of the International AAAI Conference on Web and
  Social Media}.

\bibitem[{Ingram(2016)}]{disney-salesforce}
Ingram, M. 2016.
\newblock Disney, Salesforce Dropped Twitter Bids Because of Trolls.

\bibitem[{Kumar(2019)}]{kumar2019algorithmic}
Kumar, S. 2019.
\newblock The algorithmic dance: YouTube's Adpocalypse and the gatekeeping of
  cultural content on digital platforms.
\newblock \emph{Internet Policy Review} .

\bibitem[{Lagorio-Chafkin(2018)}]{We-are-the-nerds}
Lagorio-Chafkin, C. 2018.
\newblock \emph{We Are the Nerds: The Birth and Tumultuous Life of Reddit}.
\newblock Hachette.

\bibitem[{Mami{\'e}, Ribeiro, and West()}]{mamie2021anti}
Mami{\'e}, R.; Ribeiro, M.~H.; and West, R. ????
\newblock Are Anti-Feminist Communities Gateways to the Far Right? Evidence
  from Reddit and YouTube .

\bibitem[{Mittos et~al.(2020)Mittos, Zannettou, Blackburn, and
  De~Cristofaro}]{mittos2020and}
Mittos, A.; Zannettou, S.; Blackburn, J.; and De~Cristofaro, E. 2020.
\newblock “And We Will Fight for Our Race!” A Measurement Study of Genetic
  Testing Conversations on Reddit and 4chan.
\newblock In \emph{Proceedings of the International AAAI Conference on Web and
  Social Media}.

\bibitem[{Moyer(2015)}]{Moyer-WaPo2015}
Moyer, J. 2015.
\newblock `CoonTown': A noxious, racist corner of Reddit survives recent purge.
\newblock Washington Post.

\bibitem[{Pearl(2014)}]{pearl2014interpretation}
Pearl, J. 2014.
\newblock Interpretation and identification of causal mediation.
\newblock \emph{Psychological methods} .

\bibitem[{Pieters(2017)}]{Pieters-JCR2017}
Pieters, R. 2017.
\newblock {Meaningful Mediation Analysis: Plausible Causal Inference and
  Informative Communication}.
\newblock \emph{Journal of Consumer Research} .

\bibitem[{Prabhu et~al.(2021)Prabhu, Guhathakurta, Subramanian, Reddy, Sehgal,
  Karandikar, Gulati, Arora, Shah, Kumaraguru et~al.}]{prabhu2021capitol}
Prabhu, A.; Guhathakurta, D.; Subramanian, M.; Reddy, M.; Sehgal, S.;
  Karandikar, T.; Gulati, A.; Arora, U.; Shah, R.~R.; Kumaraguru, P.; et~al.
  2021.
\newblock Capitol (Pat) riots: A comparative study of Twitter and Parler.
\newblock \emph{arXiv preprint arXiv:2101.06914} .

\bibitem[{Ribeiro et~al.(2020)Ribeiro, Blackburn, Bradlyn, De~Cristofaro,
  Stringhini, Long, Greenberg, and Zannettou}]{ribeiro2020evolution}
Ribeiro, M.~H.; Blackburn, J.; Bradlyn, B.; De~Cristofaro, E.; Stringhini, G.;
  Long, S.; Greenberg, S.; and Zannettou, S. 2020.
\newblock The Evolution of the Manosphere Across the Web.
\newblock \emph{arXiv preprint arXiv:2001.07600} .

\bibitem[{Roberts(2014)}]{roberts2014behind}
Roberts, S.~T. 2014.
\newblock \emph{Behind the screen: The hidden digital labor of commercial
  content moderation}.
\newblock Ph.D. thesis, University of Illinois at Urbana-Champaign.

\bibitem[{Scrivens, Wojciechowski, and Frank(2020)}]{scrivens2020examining}
Scrivens, R.; Wojciechowski, T.~W.; and Frank, R. 2020.
\newblock Examining the developmental pathways of online posting behavior in
  violent right-wing extremist forums.
\newblock \emph{Terrorism and Political Violence} .

\bibitem[{Simi and Futrell(2015)}]{simi2015american}
Simi, P.; and Futrell, R. 2015.
\newblock \emph{American Swastika: Inside the white power movement's hidden
  spaces of hate}.

\bibitem[{Tufekci(2012)}]{Tufekci-CITP2012}
Tufekci, Z. 2012.
\newblock If Reddit Really Regrets ``Not Taking Stronger Action Sooner'', What
  Will It Do in the Future?
\newblock Freedom to Tinker.

\bibitem[{Wohn(2019)}]{wohn2019volunteer}
Wohn, D.~Y. 2019.
\newblock Volunteer moderators in twitch micro communities: How they get
  involved, the roles they play, and the emotional labor they experience.
\newblock In \emph{Proceedings of the 2019 CHI conference on human factors in
  computing systems}.

\bibitem[{Wojcieszak(2010)}]{wojcieszak2010don}
Wojcieszak, M. 2010.
\newblock ‘Don’t talk to me’: Effects of ideologically homogeneous online
  groups and politically dissimilar offline ties on extremism.
\newblock \emph{New Media \& Society} .

\bibitem[{Wright, Trott, and Jones()}]{wright2020pussy}
Wright, S.; Trott, V.; and Jones, C. ????
\newblock ‘The pussy ain’t worth it, bro’: assessing the discourse and
  structure of MGTOW.
\newblock \emph{Information, Communication \& Society} .

\bibitem[{Zannettou et~al.(2018)Zannettou, Bradlyn, De~Cristofaro, Kwak,
  Sirivianos, Stringini, and Blackburn}]{zannettou2018gab}
Zannettou, S.; Bradlyn, B.; De~Cristofaro, E.; Kwak, H.; Sirivianos, M.;
  Stringini, G.; and Blackburn, J. 2018.
\newblock What is gab: A bastion of free speech or an alt-right echo chamber.
\newblock In \emph{Companion Proceedings of the The Web Conference 2018}.

\bibitem[{Zannettou et~al.(2020)Zannettou, ElSherief, Belding, Nilizadeh, and
  Stringhini}]{zannettou2020measuring}
Zannettou, S.; ElSherief, M.; Belding, E.; Nilizadeh, S.; and Stringhini, G.
  2020.
\newblock Measuring and characterizing hate speech on news websites.
\newblock In \emph{12th ACM Conference on Web Science}.

\end{thebibliography}

\end{document}